\documentclass{llncs}
\usepackage{float}
\usepackage{amssymb,lscape}
\usepackage{graphicx}
\usepackage{enumitem}
\usepackage{multirow}
\usepackage{hhline}
\usepackage{url}
\usepackage{bbding} 
\usepackage[table,xcdraw]{xcolor}
\usepackage{booktabs}
\usepackage{longtable}
\usepackage{ltablex}
\usepackage[normalem]{ulem}
\useunder{\uline}{\ul}{}

\urldef{\mailsa}\path|mayra.nilsson@volvocars.com|
\urldef{\mailsb}\path|vard.antinyan@volvocars.com|
\urldef{\mailsc}\path|lucas.gren@cse.gu.se|
\newcommand{\keywords}[1]{\par\addvspace\baselineskip
\noindent\keywordname\enspace\ignorespaces#1}

\begin{document}

\mainmatter  

\title{Do internal software quality tools measure validated metrics?}

\titlerunning{Do internal software quality tools measure validated metrics?}

\author{Mayra Nilson\inst{2} \and Vard Antinyan\inst{2} \and Lucas Gren\inst{1,2}
}
\authorrunning{Lucas Gren and Magdalena Lindman}

\institute{Chalmers $|$ University of Gothenburg, \\
Gothenburg, Sweden\\ 
\and
Volvo Cars,\\
Gothenburg, Sweden\\
\mailsa\\
\mailsb\\
\mailsc\\}

\toctitle{Lecture Notes in Computer Science}
\tocauthor{Authors' Instructions}
\maketitle

\begin{abstract}
Internal software quality determines the maintainability of the software product and influences the quality in use. There is a plethora of metrics which purport to measure the internal quality of software, and these metrics are offered by static software analysis tools. To date, a number of reports have assessed the validity of these metrics. No data are available, however, on whether metrics offered by the tools are somehow validated in scientific studies. The current study covers this gap by providing data on which tools and how many validated metrics are provided. The results show that a range of metrics that the tools provided do not seem to be validated in the literature and that only a small percentage of metrics are validated in the provided tools. 
\end{abstract}

\section{Introduction}
\label{intro}
Software quality has been a major concern for as long as software has existed \cite{Ref1}. Billing errors and medical fatalities can be traced to the issue of software quality \cite{Ref2}. The ISO/IEC 9126 standard defines quality as ``the totality of characteristics of an entity that bears on its ability to satisfy stated and implied needs'' \cite{Ref3}. This standard distinguishes internal and external quality. The former is the quality of pre-release software that determines the ability of a project to be maintained further. Internal quality is observed and experienced by developers. However, it affects the functionality and user experience too \cite{Ref4}. 

To assess and manage internal quality of software, internal software quality metrics are used. While much research has been conducted on such metrics in forms of empirical studies, mapping studies, and systematic literature reviews \cite{Ref6,Ref7,Ref8}, very little research has been done on the tools that implement these metrics.

One study on software metric tools concluded that there are considerable variations regarding the output from different tools for the same metric on the same software source code \cite{Ref9}. This indicates that the implementation of a given metric varies from tool to tool. We have not found any study that investigates the implemented metrics in the tools and their scientific validity in the research. We, therefore, wanted to investigate whether the metrics provided in the tools are validated in scientific studies: Studies concluding that a given metric can predict an external software quality attribute to an acceptable precision. An external attribute can for example be fault proneness, maintainability, or testability.

The value of validating an internal quality metric in regard to an external quality attributes is to provide solid measures for good predictions of external quality. For example, Santos et al.\ \cite{Ref6} conducted such a study on eight separate groups of students developing a system based on the same requirements. For each iteration of the software the metrics were studied to see if they could predict the faults that were found by independent testing.

There are many software metric tools but the validity of the metrics they provide is not investigated. Considering that the usefulness of a metric is in its validity we set out to investigate the amount of somehow validated metrics that the current popular tools provide. To the best of our knowledge there is no study which investigates the existing tools in terms of what metrics they provide. To address this research problem the following research questions were formulated:

RQ1 Which internal software quality metrics are validated in scientific studies? 

RQ2 Which are the tools that support these metrics? 

RQ3 What are the administrative capabilities of these tools to conduct measurements? 

\section{Related Work}
Internal software quality is related to the structure of the software itself as opposed to external software quality which is concerned with the behaviour of the software when it is in use. The structure of the software is not visible to the end user, but is still important since internal attributes (e.g., size, complexity, and cohesion) affect external quality attributes (e.g., maintainability and understandability) \cite{Ref11}. External quality is limited to the final stages of software development, whereas testing for internal quality is possible from the early stages of the development cycle, hence internal quality attributes have an important role to play in the improvement of software quality. The internal quality attributes are measured by means of internal quality metrics \cite{Ref12}. We, of course, recognize that the metric is not the construct in itself, but since it is assumed to be closely connected, we assume the metrics in the empirical studies are close to the construct they aim at measuring. 


Metrics can be validated for all programming languages, but some apply only to specific programming paradigms and the majority can be classified as Traditional or Object Oriented Metrics (OO)\cite{Ref16}. Considering the popularity of object oriented metrics it is not surprising that most of the validation studies concentrate on OO \cite{Ref18}. In 2012, Yeresime \cite{Ref21} performed a theoretical and empirical evaluation on a subset of the traditional metrics and object oriented metrics used to estimate a systems reliability, testing effort and complexity. The paper explored source code metrics such as cyclomatic complexity, size, comment percentage and CK Metrics (WMC, DIT, NOC, CBO, RFC LCOM). Yeresime’s studies concluded that the aforementioned traditional and object oriented metrics provide relevant information to practitioners in regard to fault prediction while at the same time provide a basis for software quality assessment.

Jabangwe et al.\ \cite{Ref22} in their systematic literature review, which focused mainly on empirical evaluations of measures, used on object oriented programs concluded that the link from metrics to reliability and maintainability across studies is the strongest for: LOC (Lines of Code), WMC McCabe (Weighted Method Count), RFC (Response for a Class) and CBO (Coupling Between Objects). Antinyan et al.\ \cite{Ref25} proved in their empirical study on complexity that complexity metrics such as McCabe cyclomatic complexity \cite{Ref26}, \cite{Ref27} measures, Fan-Out, Fan-In, Coupling Measures of \cite{Ref28}, \cite{Ref29} OO measures, Size measure \cite{Ref30} and Readability measures \cite{Ref31,Ref32} correlate strongly with maintenance time. They also suggested that more work is required to understand how software engineers can effectively use existing metrics to reduce maintenance effort. 

Another example is a recent study on client-based cohesion metrics for OO classes. The study included a multivariate regression analysis on fourteen cohesion metrics applying the backwards selection process to find the best combination of cohesion metrics that can be used together to predict testing effort, the results revealed that LCOM1  (Lack of Cohesion of Methods 1) LCOM2 (Lack of Cohesion of Methods 2), LCOM3 (Lack of Cohesion of Methods 3) and CCC (Client Class Cohesion) are significant predictors for testing effort in classes \cite{Ref33}.



\section{Research Method}
First, a review of previous work on software metrics validation was done. The main goal was to elicit internal quality measures based on their existence in scientific studies. We only consider empirical validation in this study and therefore exclude theoretical validation without industry, or ``real,'' data. As a second step, we searched and selected tools that support these metrics. Afterwards, relevant data was investigated about the tools which could help with some practicalities when using these tools in software development practice. Finally, consistency of measurement was evaluated between these tools on a set of open source projects.

\subsection{Search and Identification of Relevant Papers and Metrics}
To identify relevant research papers, the subject area was restricted to Engineering and Computer Science, the string below was built based on keywords as well as synonyms defined for the study. We used the folowing digital libraries and search engines: Google Scholars\footnote[2]{https://scholar.google.se/}, IEEE Digital Library\footnote[3]{http://ieeexplore.ieee.org}, Science Direct\footnote[4]{http://www.sciencedirect.com}, Springer\footnote[5]{http://www.springer.com} and Engineering Village\footnote[6]{https://www.engineeringvillage.com}. 

The following search string was used: 
\begin{center}
\textit{(validated OR ``verification of internal quality'' OR ``code quality'' OR ``software quality'') AND (metric OR metrics OR measure OR measuring) AND tools}
\end{center}

After the search 567 articles were found (Fig. 1) many of which were irrelevant for the purpose of this paper.

\begin{center}
\begin{figure}[ht]
 \includegraphics[scale=0.3] {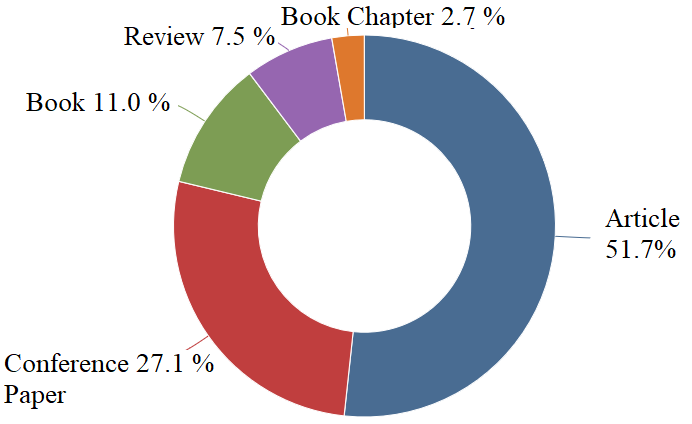}
\caption{Pie chart showing types and percentage of papers found}
\label{fig:1}       
\end{figure}
\end{center}

These papers were assessed according the criteria described in Table~\ref{tab:1}. The output from this step resulted in 292 research papers subjecting 30 internal quality metrics to validation. Afterwards, from these metrics we selected such metrics that were considered somehow validated as results of the papers. If any metric was evaluated inconsistently in different papers we excluded it from our final list. We realize the issue with this type of exclusion criteria, but we wanted an initial assessment of the existence of somehow validated metrics in the available tools. The main criteria of validation was that a metric shall have tangible correlation with an external software quality attribute such as maintainability and defects. However, since the aim of the study was code improvement metrics particularly, we added one more criteria: Besides having tangible correlation, a metric shall also be possible to manipulate for influencing the external quality attributes, see e.g.\ \cite{Ref41}, that is increasing maintainability or decreasing defects. For example, Lines of Code usually has a tangible correlation with defects and maintainability. However, Lines of code is essential for writing code, and therefore cannot be reduced for the purpose of decreasing defects or increasing maintainability. Similarly, McCabe's cyclomatic complexity has tangible correlation with defect and maintainability but cannot be manipulated to a significant degree for code improvements. It should be clear, however, that if a metric is not supported for quality assessment in a validation study it can be valid for other activities. For example, the same Lines of Code metric can be supported for maintenance effort estimation. Or Cyclomatic complexity can be useful for testability assessment.

\begin{table}[h]
\caption{Criteria}
\label{tab:1}       
\begin{tabular}{ll}
\hline
\multicolumn{2}{l}{\textit{\textbf{Inclusion Criteria}}}                                                                                                                       \\ \hline
\textbf{I1} & Papers published in a well-known software engineering journals or conferences                                                                                                                      \\ \hline
\textbf{I2} & \begin{tabular}[c]{@{}l@{}}Papers that present studies in empirical validation of internal \\ quality or software metrics\end{tabular}           \\ \hline
\multicolumn{2}{l}{\textit{\textbf{Exclusion Criteria}}}                                                                                                                       \\ \hline
\textbf{E1} & Papers that are not written in English                                                                                                                           \\ \hline
\textbf{E2} & \begin{tabular}[c]{@{}l@{}}Papers that do not have internal metrics context and do not provide scientific \\ validation of internal quality metrics\end{tabular} \\ \hline
\end{tabular}
\end{table}

\subsection{Selection of Tools}
For the selection of the tools, first, a free search on the Internet was conducted. The main criteria was that the tools should conduct any type of static analysis. As a result 130 tools were found. Because the aim of this paper is to aid practitioners to improve the quality of their code, and because there are several tools that support the same metrics, we set criteria for the selection of the tools. A summary of the criteria is presented in Table~\ref{tab:2}.

\begin{table}[ht]
\caption{Criteria for Tool Selection}
\label{tab:2}       
\begin{tabular}{ll}
\hline
\textbf{Criteria} &                                                                                                                                              \\ \hline
C1                & Support automated static analysis                                                                                                                      \\ \hline
C2                & Offer at least one somehow validated metric                                                                                                                \\ \hline
C3                & Integration to IDEs and version control systems                                                                                                                \\ \hline
C4                & Be free or at least offer a trial option                                                                                                            \\ \hline
C5                & Support at least two programming languages                                                                                                     \\ \hline

C6                & \begin{tabular}[c]{@{}l@{}} Provide documentation such as user manual and/or installation manual\end{tabular}                                                                          \\ \hline
\end{tabular}
\end{table}

\section{Results}
\subsection{Selection of Metrics}
The first research question was what are the somehow validated metrics according to literature. Based on the inclusion and exclusion criteria described earlier a total number of 292 papers were found which evaluate internal quality metrics. After an in-depth analysis of each of the paper a preliminary table with 30 metrics was created (Table \ref{tab:3}). This table presents all metrics that had been subjected to validation in an empirical study. 

\begin{table}[ht]
\caption{List of metrics subjected to validation in literature and the ones we selected for this study.}
\label{tab:3}     
\scalebox{0.8}{
\begin{tabular}{|c|l|c|c|}
\hline
\rowcolor[HTML]{C0C0C0} 
\textbf{\#} & \textbf{Metric}                                             & \textbf{No of Papers} & \textbf{Selected} \\ \hline
1           & Lack of Cohesion on Methods                                 & 9            &     X    \\ \hline
2           & Weight Methods per Class                                    & 9             &        \\ \hline
3           & Depth of Inheritance                                        & 8              &   X    \\ \hline
4           & Response For Classes                                        & 8               &  X    \\ \hline
5           & Number Of Classes                                           & 8                &     \\ \hline
6           & Coupling Between Objects                                    & 7                 &  X  \\ \hline
7           & Tight Class Cohesion                                        & 6                  & X  \\ \hline
8           & Loose Class Cohesion                                        & 5   &         X         \\ \hline
9           & Lines of Code                                               & 4    &                 \\ \hline
10          & McCabe Complexity                                           & 4     &                \\ \hline
11          & Lack of Cohesion on Methods 2                               & 4      &           X    \\ \hline
12          & Lack of Cohesion on Methods 3                               & 3       &           X   \\ \hline
13          & Lack of Cohesion on Methods 1                               & 3        &           X  \\ \hline
14          & Degree of Cohesion (Direct)                                 & 3         &            \\ \hline
15          & Degree of Cohesion (Indirect)                               & 3          &           \\ \hline
16          & Fan-Out Fan-In                                               & 2           &       X   \\ \hline
17          & Number of Methods                                           & 2            &         \\ \hline
18          & Block depth                                                 & 2             &      X  \\ \hline
19          & Weight Methods per Class-McCabe                             & 1              &       \\ \hline
20          & Standard Deviation Method Complexity                        & 1               &      \\ \hline
21          & Average Method Complexity                                   & 1                &     \\ \hline
22          & Maximum Cyclomatic Complexity of a single Method of a Class & 1                 &    \\ \hline
23          & Number of Instance Methods                                  & 1   &                  \\ \hline
24          & Number of Trivial Methods                                   & 1    &                 \\ \hline
25          & Number of send Statements defined in a Class                & 1     &                \\ \hline
26          & Number of ADT Defined in A Class                            & 1      &               \\ \hline
27          & Sensitive Class Cohesion                                    & 1       &              \\ \hline
28          & Improved Connection Based on Member Connectivity            & 1        &             \\ \hline
29          & Lack of Cohesion on Methods 4                               & 1         &    X        \\ \hline
30          & Number of Attributes                                        & 1          &           \\ \hline

\end{tabular}}
\end{table}

In the next step all the metrics that were evaluated as somehow validated were selected.  After excluding metrics that are invalid for internal quality assessment the final list of 12 validated metrics also presented in Table \ref{tab:3}. These are the metrics that were consistently shown to have tangible correlation with quality attributes such as maintainability and defects. At the same time, these are the metrics that can be useful when improving the code, due to the fact that they are correlated to external quality attributes, and possible then also has an effect on them. We recognize here that correlation is not causation but the causality challenge in software engineering research is a much larger discussion, and we assume a potential causality here even if that might not be true with a lot of confounding factors in practice. it would at least be a useful start for practitioners when improving code, we argue.

\subsection{The Selected Tools}

The research question 2 concerns the administrative capabilities of the tools. In total there are over 130 commercial and non-commercial tools that claim to provide internal software quality metrics. Out of these 130 only six tools were selected according to the selection criteria. Interestingly, these tools also represent some of the widely known tools that researchers and software development organizations have often reported to use. Brief descriptions of the tools can be found on their respective homepages: QAC\footnote[7]{https://www.qa-systems.com}, Understand\footnote[8]{https://www.scitools.com}, CPPDepend\footnote[9]{https://www.cppdepend.com}, SonarQube\footnote[11]{https://www.sonarqube.com}, 
Eclipse Metrics Plugin\footnote[12]{eclipse-metrics-sourceforge.net}, and SourceMonitor\footnote[14]{https://www.campwoodsw.com}.

An additional difficulty of the tool selection was that most of the commercial tools' trial versions did not allow for a proper evaluation, meaning that, for example, reports generated by the tools were unable to save, print or export, the access to all metrics or features supported were not available for free trials. Moreover, some of them required legal binding contracts for the trial as well as written clarification of the purpose and the context in which the tool´s reports will be used. 

The third research question was about the administrative capabilities of the tools. Table \ref{tab:7} shows whether a given tool has a given capability for all the tools and all the defined capabilities. If a tool has the given capability the intersection cell is marked with ``1'', otherwise ``0'' is marked. A total score at the bottom of the table shows the sum of all marks.

The second research question is concerned with which tools support the validated metrics. No tool was found that supports all of the metrics in Table 5, which indicated that it is reasonable to select tools that support as many of the validated metrics as possible. A preliminary search of the prospects for each tool indicated that several somehow validated metrics were supported, but a deeper analysis of the technical documentation showed that this was not always the case, since some metrics either did not exist in the tool or existed but under a different name than in Table 5. In the latter cases it was unclear whether the metric was implemented according to the design of the original scientific paper or underwent modifications. These metrics were included as valid metric implementation, however, since the overall implementations of validated metrics in the tools were scarce. 

Table 4 presents all six tools, the amount of all metrics that they offer and the somehow validated metrics that they offer. The second column of the table shows the number of all metrics versus the number of somehow validated metrics that each of the tools provide. In the third column the validated metrics names are registered. Generally it appears that tools provide substantially larger amount of metrics when compared to the number of validated metrics. Moreover, the tools provide only half or less of validated metrics. This point is crucial and is discussed in great detail in the discussion below. 

\begin{table}[H]
\caption{All the somehow validated metrics offered by the tools.}
\label{tab:7}
\begin{tabular}{|l|c|l|}
\hline
\multicolumn{1}{|c|}{\textbf{Tool}} & \textbf{All / Validated} & \multicolumn{1}{c|}{\textbf{validated metrics}}                                  \\ \hline

QA-C                    & 66 / 1    &    nesting depth   \\ \hline              Understand              & 102 / 6                         & \begin{tabular}[c]{@{}l@{}}fan-in, fan-out, depth of inheritance, \\
response for classes,  coupling  between objects, \\
lack of cohesion between methods\end{tabular} \\ \hline
CPPDepend               & 40 / 2    & depth of inheritance, lack of cohesion on methods \\ \hline
SonarQube               & 59 / 0    &            \\ \hline
Eclipse m. plugin  & 28 / 2    & depth of inheritance, lack of cohesion on methods \\ \hline
SourceMonitor           & 12 / 2    & depth of inheritance, nesting depth           \\ \hline
\end{tabular}
\end{table} 

\section{Discussion}
First, it should be clear that the tools evaluated in this study may provide other useful functionalities besides measurements. For example, some of the tools can provide code violation detection. The scope of this paper is only concerned with the metrics they provide.
The results show that there are three important points concerning the first research question. The first point is that there were a number of metrics that were designed theoretically and reported in scientific papers, which purported to be useful as internal software quality measures. These are 30 metrics presented in Table 3.

The second point is that as a first indication of validity these metrics shall be able to predict some external software quality attributes, such as maintainability, testability, defect-proneness, etc. For example Radjenović [44] found 106 publications which evaluated one or another set of these metrics against defect prediction.

The third and final point is that being a good predictor does not entail being useful quality measure. An additional criteria for validity is whether the measure can be manipulated for making code quality better. For example Lines of code measure is a good predictor of defects, but it is not possible to manipulate lines of code to reduce the defects, because lines of code is an essential element of creating code \cite{Ref43}. Considering the last two points we got twelve somehow validated metrics (marked with an ``X'' in Table 3).

\begin{table}[H]
\caption{Tool characteristics and scores}
\label{tab:7}
\scalebox{0.8}{
\begin{tabular}{|l|c|c|c|c|c|c|c|}
\hline
\rowcolor[HTML]{C0C0C0} 
\multicolumn{1}{|c|}{\cellcolor[HTML]{C0C0C0}\textbf{\begin{tabular}[c]{@{}c@{}}Tool \\ characte-\\ ristics\end{tabular}}}   & \textbf{\begin{tabular}[c]{@{}c@{}}Tool \\ sub\\ characte-\\ ristics\end{tabular}} & \textbf{QAC} & \textbf{\begin{tabular}[c]{@{}c@{}}Under-\\ stand\end{tabular}} & \textbf{\begin{tabular}[c]{@{}c@{}}CPP\\ Depend\end{tabular}} & \textbf{\begin{tabular}[c]{@{}c@{}}Sonar\\ Qube\end{tabular}} & \textbf{\begin{tabular}[c]{@{}c@{}}Eclipse \\ Metrics \\ Plug-in\end{tabular}} & \textbf{\begin{tabular}[c]{@{}c@{}}Source\\ Monitor\end{tabular}} \\ \hline
\multicolumn{1}{|c|}{\textbf{Language}}                                                                                      & \begin{tabular}[c]{@{}c@{}}Java \\ (Android)\end{tabular}                          & 0            & 1                                                               & 1                                                             & 1                                                             & 1                                                                              & 1                                                                 \\ \hline
\textbf{}                                                                                                                    & C,C++                                                                              & 1            & 1                                                               & 1                                                             & 1                                                             & 0                                                                              & 1                                                                 \\ \hline
\textbf{}                                                                                                                    & Phyton                                                                             & 0            & 1                                                               & 0                                                             & 1                                                             & 0                                                                              & 0                                                                 \\ \hline
\textbf{}                                                                                                                    & Csharp                                                                                 & 0            & 1                                                               & 1                                                             & 1                                                             & 0                                                                              & 1                                                                 \\ \hline
\textbf{}                                                                                                                    & \begin{tabular}[c]{@{}c@{}}Delphi/\\ Pascal\end{tabular}                           & 0            & 1                                                               & 0                                                             & 0                                                             & 0                                                                              & 1                                                                 \\ \hline
\textbf{}                                                                                                                    & \begin{tabular}[c]{@{}c@{}}Visual\\ Basic\end{tabular}                             & 0            & 1                                                               & 0                                                             & 1                                                             & 0                                                                              & 1                                                                 \\ \hline
\textbf{}                                                                                                                    & \begin{tabular}[c]{@{}c@{}}HTML/\\  .NET\end{tabular}                              & 0            & 1                                                               & 1                                                             & 1                                                             & 0                                                                              & 1                                                                 \\ \hline
\rowcolor[HTML]{EFEFEF} 
\multicolumn{1}{|c|}{\cellcolor[HTML]{EFEFEF}\textbf{Availability}}                                                          & \begin{tabular}[c]{@{}c@{}}Open \\ Source\end{tabular}                             & 0            & 0                                                               & 0                                                             & 1                                                             & 1                                                                              & 0                                                                 \\ \hline
\rowcolor[HTML]{EFEFEF} 
                                                                                                                             & Free                                                                               & 0            & 0                                                               & 0                                                             & 0                                                             & 0                                                                              & 1                                                                 \\ \hline
\rowcolor[HTML]{EFEFEF} 
                                                                                                                             & Free-Trial                                                                         & 0            & 1                                                               & 1                                                             & 0                                                             & 0                                                                              & 0                                                                 \\ \hline
\rowcolor[HTML]{EFEFEF} 
                                                                                                                             & Commercial                                                                         & 1            & 1                                                               & 1                                                             & 1                                                             & 0                                                                              & 0                                                                 \\ \hline
\multicolumn{1}{|c|}{\textbf{Interface}}                                                                                     & \begin{tabular}[c]{@{}c@{}}Dynamic\\  GUI\end{tabular}                             & 1            & 1                                                               & 1                                                             & 1                                                             & 1                                                                              & 1                                                                 \\ \hline
                                                                                                                             & \begin{tabular}[c]{@{}c@{}}Export \\ Graphs\end{tabular}                           & 1            & 1                                                               & 1                                                             & 1                                                             & 0                                                                              & 1                                                                 \\ \hline
                                                                                                                             & \begin{tabular}[c]{@{}c@{}}Customized \\ Metrics \\ Tables\end{tabular}            & 1            & 1                                                               & 1                                                             & 0                                                             & 0                                                                              & 0                                                                 \\ \hline
                                                                                                                             & \begin{tabular}[c]{@{}c@{}}Metrics \\ Automation \\ via CMD\end{tabular}           & 1            & 0                                                               & 0                                                             & 0                                                             & 0                                                                              & 0                                                                 \\ \hline
\rowcolor[HTML]{EFEFEF} 
\multicolumn{1}{|c|}{\cellcolor[HTML]{EFEFEF}\textbf{\begin{tabular}[c]{@{}c@{}}Supported\\  OS\end{tabular}}}               & Windows                                                                            & 1            & 1                                                               & 1                                                             & 1                                                             & 1                                                                              & 1                                                                 \\ \hline
\rowcolor[HTML]{EFEFEF} 
                                                                                                                             & Mac                                                                                & 1            & 1                                                               & 1                                                             & 1                                                             & 1                                                                              & 1                                                                 \\ \hline
\rowcolor[HTML]{EFEFEF} 
                                                                                                                             & Ubuntu                                                                             & 1            & 1                                                               & 1                                                             & 1                                                             & 1                                                                              & 1                                                                 \\ \hline
\rowcolor[HTML]{EFEFEF} 
                                                                                                                             & Cloud                                                                              & 1            & 1                                                               & 1                                                             & 0                                                             & 0                                                                              & 0                                                                 \\ \hline
\multicolumn{1}{|c|}{\textbf{Integration}}                                                                                   & IDE                                                                                & 1            & 1                                                               & 1                                                             & 1                                                             & 1                                                                              & 0                                                                 \\ \hline
                                                                                                                             & \begin{tabular}[c]{@{}c@{}}Continuous \\ Integration \\ Tools\end{tabular}         & 1            & 1                                                               & 1                                                             & 1                                                             & 0                                                                              & 0                                                                 \\ \hline
                                                                                                                             & \begin{tabular}[c]{@{}c@{}}Version \\ Control\\ Tools\end{tabular}                 & 1            & 0                                                               & 0                                                             & 1                                                             & 0                                                                              & 0                                                                 \\ \hline
                                                                                                                             & \begin{tabular}[c]{@{}c@{}}Issue \\ Tracker \\ Tools\end{tabular}                  & 1            & 0                                                               & 0                                                             & 1                                                             & 0                                                                              & 0                                                                 \\ \hline
\rowcolor[HTML]{EFEFEF} 
\multicolumn{1}{|c|}{\cellcolor[HTML]{EFEFEF}\textbf{\begin{tabular}[c]{@{}c@{}}Compliance \\ to \\ standards\end{tabular}}} & MISRA                                                                              & 1            & 1                                                               & 0                                                             & 0                                                             & 0                                                                              & 0                                                                 \\ \hline
\rowcolor[HTML]{EFEFEF} 
\multicolumn{1}{|c|}{\cellcolor[HTML]{EFEFEF}}                                                                               & \begin{tabular}[c]{@{}c@{}}ISO \\ 26262\end{tabular}                               & 1            & 0                                                               & 0                                                             & 0                                                             & 0                                                                              & 0                                                                 \\ \hline
\rowcolor[HTML]{EFEFEF} 
\multicolumn{1}{|c|}{\cellcolor[HTML]{EFEFEF}}                                                                               & CWE                                                                                & 1            & 0                                                               & 0                                                             & 0                                                             & 0                                                                              & 0                                                                 \\ \hline
\rowcolor[HTML]{EFEFEF} 
\multicolumn{1}{|c|}{\cellcolor[HTML]{EFEFEF}}                                                                               & CERT                                                                               & 1            & 0                                                               & 0                                                             & 0                                                             & 0                                                                              & 0                                                                 \\ \hline
\rowcolor[HTML]{EFEFEF} 
\multicolumn{1}{|c|}{\cellcolor[HTML]{EFEFEF}}                                                                               & \begin{tabular}[c]{@{}c@{}}ISO/IEC \\ 9899:2011\end{tabular}                       & 0            & 0                                                               & 0                                                             & 0                                                             & 0                                                                              & 0                                                                 \\ \hline
\multicolumn{1}{|c|}{\textbf{Documentation}}                                                                                 & Yes                                                                                & 1            & 1                                                               & 1                                                             & 1                                                             & 1                                                                              & 1                                                                 \\ \hline
                                                                                                                             & Tutorials                                                                          & 1            & 1                                                               & 1                                                             & 1                                                             & 0                                                                              & 0                                                                 \\ \hline
\multicolumn{1}{|c|}{\textbf{\begin{tabular}[c]{@{}c@{}}Metric\\ resolution\end{tabular}}}                                   & Project                                                                            & 1            & 1                                                               & 1                                                             & 1                                                             & 1                                                                              & 1                                                                 \\ \hline
                                                                                                                             & File                                                                               & 1            & 1                                                               & 0                                                             & 1                                                             & 0                                                                              & 0                                                                 \\ \hline
                                                                                                                             & \begin{tabular}[c]{@{}c@{}}Function/\\ Method\end{tabular}                         & 1            & 1                                                               & 0                                                             & 0                                                             & 0                                                                              & 1                                                                 \\ \hline
\rowcolor[HTML]{C0C0C0} 
\multicolumn{1}{|c|}{\cellcolor[HTML]{C0C0C0}\textbf{\begin{tabular}[c]{@{}c@{}}Total Score in \\ Percentage\end{tabular}}}  & \multicolumn{1}{l|}{\cellcolor[HTML]{C0C0C0}}                                      & \textbf{23}  & \textbf{24}                                                     & \textbf{18}                                                   & \textbf{21}                                                   & \textbf{9}                                                                    & \textbf{15}                                                       \\ \hline
\end{tabular}}
\end{table}

Research question 2 concerns with finding tools for supporting these twelve metrics. Unfortunately the currently available tools poorly capture the validated metrics. Moreover, they provide an excessive number of other metrics, the use of which are not clear. This confusion is deepened more when it turns out that there are several ways for measuring the same metric. For example, when counting function calls (fan-out) it is not clear whether unique functions calls or all function calls should be counted. Furthermore, the same function can be called multiple times with different argument list. Similarly "switch --- case" problem for cyclomatic complexity, or "commented --- non-commented lines" for lines of code. 

Research question 3 is concerned by capabilities of the tools. Most of the tools had specific technical requirements for the code to be analyzed. For example, some tools were not able to start the analysis without a compiled file. Several tools required a specific hardware in order to use their servers to run the static code analysis, however none of these are explicitly mentioned in their documentation. 
There are also big differences between the commercial and free tools, where the commercial tools offer a sometimes overwhelming level of detail and the free tools can be somewhat superficial.

Another issue with the tools is that they do not always support reporting results on the same level. The measurements can be reported on entities such as project, file or function/method level, but not all tools support these levels. The most relevant level would normally be function/method level, since this level can be assigned to a developer or a team for improvements. 

Table 5 summarizes capabilities of the tools on a rough level. But much more details reveal when using the tools. One clear point is that only a portion of the validated metrics are actually supported in any of the tools. It is striking that the tools offer overwhelming numbers of metrics (if they can be called metrics), as though every additional metric provides some value (Table 4). But they are likely to provide confusion, because simplistic and not validated metrics risk their usefulness.

\section{Threats of Validity}

There could be errors in the underlying studies, and the validity of the metrics is established in those other papers. The results from those validation studies could be incorrect, which could influence the result of this study. The threat to validity for a specific metric can be assumed to be lower the more independent validation studies have been conducted. This threat was somewhat mitigated by the fact that the selected metrics are supported by 2 or more papers. 

Another threat is a potential error in the search process. This kind of limitation is particularly difficult to tackle, we tried to use a not too broad and not to narrow search string to capture relevant papers. 

Omission of relevant papers is a third threat we have identified, and as stated in the research method section, during the initial search 567 papers were found but many of these were not relevant to this study as they also included papers about medicine, biochemistry, environmental science, chemistry, agriculture, physics or social science. The reason for these papers being found by the search is presumably that the keywords ``metrics,'' ``software'' and ``validation'' are common to many scientific papers. In the second search the subject areas above were excluded and as a result 292 papers were found. There is a chance that there is an unidentified paper conducting metric validation, however, this will not change the general ratio of generally offered and somehow validated metrics. 

Naming of metrics is an external threat to validity. Unfortunately each tool can use different names for metrics in scientific papers, which could lead to a mapping problem.  

\section{Conclusions}
This study found 12 somehow validated metrics that can be helpful in practice for code improvements. Popular static code analysis tools capture these metrics partially. But at the same time, they provide overwhelming number of other metrics, the purpose of which are not clear. This may cause confusion for practitioners. Additionally, these tools have capabilities for being integrated to software development environment and helping developers in continuous code improvements, even though they are oriented to specific programming languages.

\bibliographystyle{splncs.bst}
\bibliography{refs}
\newpage

\end{document}